\documentclass{IEEEtran}
\usepackage[breaklinks]{hyperref}
\usepackage{color, dsfont, bm, epsfig, graphicx, cite, multirow, multicol, comment}
\usepackage[ruled]{algorithm2e}
\usepackage{amsmath,amsfonts,amsthm,amssymb,subfigure}
\usepackage{stfloats,booktabs,threeparttable, tabularx}
\usepackage{mathrsfs}

\newcommand{\st}[1]{\text{s.t.}}

\usepackage{xcolor}

\psfull
\begin{document} 
\title{Edge Large AI Models: Revolutionizing 6G Networks}
\author{
	\normalsize{Zixin Wang, Yuanming Shi, Yong Zhou, Jingyang Zhu, and Khaled. B. Letaief}
 %    	\thanks{Z. Wang and K. B. Letaief are with the Department of Electronic and Computer Engineering, The Hong Kong University of Science and Technology, Hong Kong (E-mail: \{eewangzx, eekhaled\}@ust.hk). Y. Shi, Y. Zhou, and J. Zhu are with the School of Information Science and Technology, ShanghaiTech University, Shanghai, 201210, China (E-mail: \{shiym, zhouyong, zhujy2\}@shanghaitech.edu.cn).
	% }
	\thanks{Z. Wang and K. B. Letaief are with The Hong Kong University of Science and Technology, Hong Kong. Y. Shi, Y. Zhou, and J. Zhu are with ShanghaiTech University, China.
	}
}	
{}
\maketitle{
\begin{abstract}
Large artificial intelligence models (LAMs) possess human-like abilities to solve a wide range of real-world problems, exemplifying the potential of experts in various domains and modalities. 
By leveraging the communication and computation capabilities of geographically dispersed edge devices, edge LAM emerges as an enabling technology to empower the delivery of various real-time intelligent services in 6G. 
Unlike traditional edge artificial intelligence (AI) that primarily supports a single task using small models, edge LAM is featured by the need of the decomposition and distributed deployment of large models, and the ability to support highly generalized and diverse tasks. 
However, due to limited communication, computation, and storage resources over wireless networks, the vast number of trainable neurons and the substantial communication overhead pose a formidable hurdle to the practical deployment of edge LAMs. 
In this paper, we investigate the opportunities and challenges of edge LAMs from the perspectives of model decomposition and resource management. 
Specifically, we propose collaborative fine-tuning and full-parameter training frameworks, alongside a microservice-assisted inference architecture, to enhance the deployment of edge LAM over wireless networks. Additionally, we investigate the application of edge LAM in air-interface designs, focusing on channel prediction and beamforming. These innovative frameworks and applications offer valuable insights and solutions for advancing 6G technology. 

%为了获得更快的服务延迟，在网络边缘端部署大模型引发了大家的兴趣。
%然而在边缘段受限的通算存环境下，海量的神经元和其之间巨额的通信量对带来了巨大的挑战。
%相比于传统的边缘AI应用，主要以单一任务的小型模型，边缘大模型的关键在于需要对模型、任务进行分解和分布式部署，以实现泛化性强，多样化的功能。
%因此，本文提出了用于训练和推理阶段的边缘段大模型的新分解架构，并结合本地无线资源进行配套的资源调度设计。进一步，我们将分布式边缘大模型应用于无线空口设计，针对信道预测，波束赋形和端到端资源优化，以提高频谱效率
\end{abstract}
}
% \begin{IEEEkeywords}
% Foun
% \end{IEEEkeywords}

\section{Introduction}
% Witnessing the remarkable achievements in large artificial intelligence model (LAM) has being revolutionizing various fields from natural language processing (NLP) to a wide range of real-world complex tasks.

With the remarkable advancement in artificial intelligence (AI), large AI models (LAMs) now excel at performing real-world complex tasks. 
They are cross-modal, highly generalizable, and adept at knowledge transfer, showcasing the vast potential of artificial general intelligence \cite{10398474}. 
With revolutionary model architectures, enormous parameter sizes, abundant data, and substantial computational resources, LAMs gain unprecedented generalization capabilities from pre-training. 
They can be adapted to a wide range of downstream tasks with only a few-shot or even zero-shot learning while achieving comparable performance to human intelligence.
Such capabilities open up groundbreaking applications across various domains, promising a new era of omnipotent intelligence. 
For instance,  92\% Fortune 500 companies have adopted ChatGPT as part of their operational infrastructure, while vehicles equipped with Tesla's FSD have already traveled over 2.575 billion kilometers. 
In addition, training a LAM with high-quality telecom-specific datasets unlocks its potential to effectively tackle various tasks across telecom domains \cite{10384630}. 
However, many emerging applications (e.g., autonomous driving and robots) are supported by cloud servers, inevitably leading to severe privacy concerns and high-latency decision-making. 
Thus, pushing LAMs with strong generalization capabilities to the network edge in 6G empowers the delivery of personalized, real-time, trustworthy intelligent services to mobile users. 

% 第一句话之后开始讲边缘大模型的训练要干什么，给个定义；However, 传统边缘AI训练要rely on federated learning，每个设备要训练完整的模型；
% 边缘大模型的推理要干什么，给个定义（多模态？），边缘AI推理处理单模态数据，特征提取压缩什么的。。。
% Edge LAMs provide many new opportunities for constructing 6G mobile networks with built-in AI.
However, the challenges faced by edge LAM are also evident. The main characteristics of LAM can be summarized as three ``gigantic" aspects: a massive number of parameters, extensive data supply, and substantial computational demands.
For example, GPT-3 has 170 billion parameters and is trained on tens of thousands of Nvidia V100 GPUs, utilizing approximately 300 billion words and 570 GB of training data.
Conventional edge AI training relies on federated learning (FL), where the entire model is trained on resource-constrained edge devices \cite{Tao2025Federated}. This approach is infeasible for edge LAMs training due to the limited computation, storage, and communication resources of edge devices.
For edge LAM inference, multiple downstream tasks require handling multimodal input data, which must sequentially pass through various LAM modules to produce inference results. 
However, traditional edge AI inference is limited to handling single-task and single-modality scenarios, which is insufficient to meet the multimodal multi-task demands.
{\color{black}
Given the above challenges inherent in edge LAM deployment, addressing these issues necessitates the establishment of distributed and trustworthy training frameworks, the assurance of resource-efficient and low-latency inference architecture, and the exploration of its application in air-interface design, to fully harness the potential of edge LAM in revolutionizing 6G networks.
% To overcome this hurdle, the key to the deployment of edge LAM lies in the decomposition of both models and tasks, followed by the allocation of scarce computation, communication, and storage resources, including the following perspectives.

%第一句将做的事情的定义，讲相比传统edge ai不同与挑战。所以我们要做什么事情，在这篇文章里我们提出了什么，这个方案解决了什么问题，有什么好处？具体针对这个方案里面我们需要做什么？
\textbf{Training}: The training of edge LAM involves both fine-tuning and learning from scratch (e.g., full-parameter training) using sensitive data on distributed edge devices, aiming to improve learning performance and enhance trustworthiness.}
This process involves updating over at least several hundred million parameters, which is unaffordable for a single edge device, necessitating the decomposition of the learning model across devices and servers. 
Thus, various parallelization frameworks, e.g., data, tensor, and pipeline parallelism \cite{deng2024cloud}, can be adopted for the decomposition, facilitating real-time and collaborative training at the network edge. 
% However, these resultant AI traffic flows exhibit characteristics such as low entropy, burstiness, and high peak utilization, posing significant challenges to resource-constrained edge networks.
However, existing parallelization frameworks are primarily designed for cloud computing centers and centralized datasets, where data privacy is not a primary concern, unlike in edge LAM training. 
% However, privacy issues are paramount in edge LAM training, especially when dealing with sensitive domain-specific data. 
Hence, it is crucial to enhance the existing parallelization frameworks for edge LAMs to safeguard local data privacy and establish effective collaborative mechanisms while accounting for diverse requirements of computation, communication, and storage resources of different parallelization frameworks. 
% Thus, we shall present a federated fine-tuning (FedFT) framework over wireless networks in Section \ref{sec: training}, which involves novel model partitioning, resource allocation, and synchronization schemes to support efficient edge LAM training \cite{10855336}. 
\begin{figure*}[t]
	\centering
	\includegraphics[width=\linewidth]{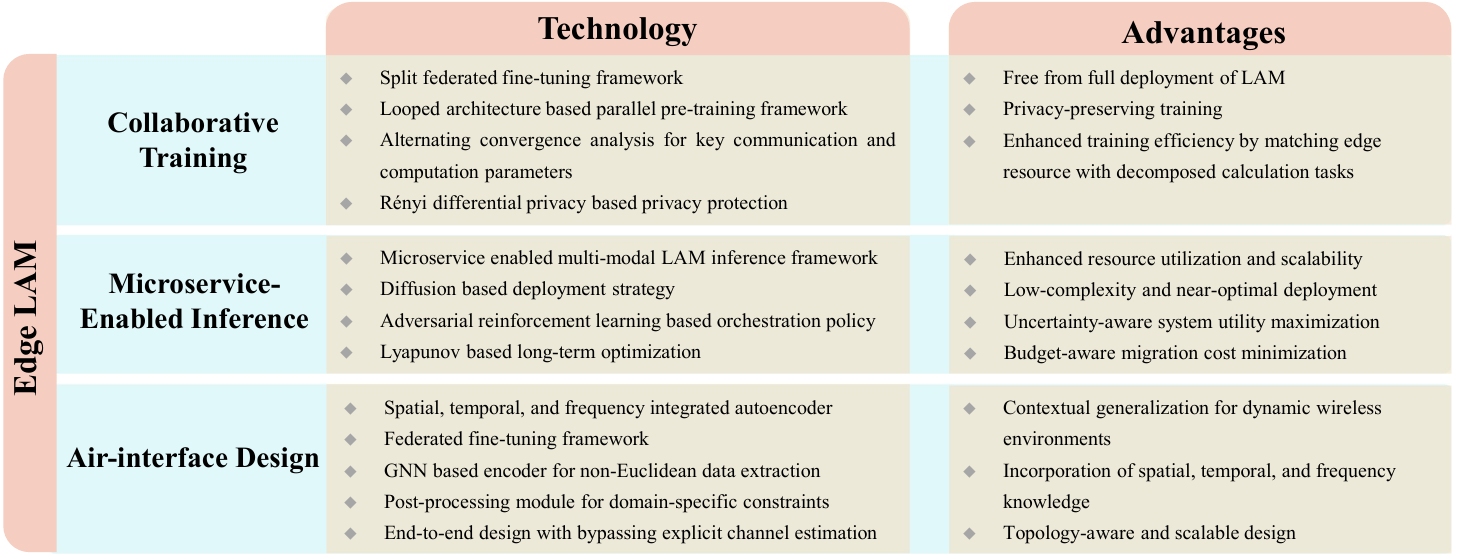}
	\caption{{\color{black}Summary of the techniques and advantages in the proposed edge LAMs framework.}}
	\vspace{-0.5cm}
	\label{fig:system_sum}
\end{figure*}

{\color{black}\textbf{Inference}: Edge LAM inference refers to providing real-time intelligent services for various
applications with different modalities requested from users, thereby improving the overall inference
timeliness and performance.}
These multimodal applications, such as autonomous driving and embodied AI, involve the sequential processing of user requests through a series of well-trained neural networks.
However, unlike inference tasks with single modality in traditional edge AI, these processes face significant challenges such as redundant computation and low resource utilization due to repeated module deployments at the network edge. 
% For instance, a single roadside unit may deploy multiple multimodal LAMs to address various traffic management needs, where LAMs utilize the same visual modality, resulting in redundant video encoding modules for environmental information collection. 
This redundancy occupies limited edge resources during invocation and increases the overall inference latency. 
% With this observation, we in Section \ref{sec: air-interface} shall present a microservice-based edge LAM inference framework that decomposes LAM into functional modules deployed at edge devices, reorganizing the multimodal downstream inference services as dynamic microservice flows. 
% The key to reducing inference latency is aligning network resources with service demands, despite their stochastic and dynamic nature \cite{yang2024latency}. This article further explores microservice deployment, robust orchestration, and online migration for edge LAM inference through real-time learning and optimization.	

{\color{black}\textbf{Application}: AI is a pivotal tool in the design and optimization of air-interface technologies, facilitating the development of intelligent and efficient wireless systems \cite{letaief2019roadmap}.}
Despite recent progress, existing AI-based methods are typically designed to tackle specific problems in specific scenarios. 
% In particular, with constrained model size and limited training data, existing AI-based methods often struggle with poor generalization, limiting their ability to effectively perform various tasks in diverse scenarios. 
{\color{black}In particular, with constrained model size and limited training data, existing AI-based methods often struggle with poor generalization, limiting their multi-tasking ability in diverse scenarios. }
This issue is further exacerbated in highly dynamic wireless networks, where frequent retraining is required, presenting significant challenges in terms of communication overhead, computation costs, and achievable performance. 
LAMs with billions of parameters exhibit unprecedented levels of general intelligence and can be fine-tuned to simultaneously support diverse network management and optimization tasks \cite{10778660, 10.1145/3651890.3672268}. 
Despite significant progress, it is still challenging to apply LAM to the design and optimization of air-interface technologies, as the complexity of wireless channels, intricate network architecture, and the need for low-latency decision-making should be considered. 
% Thus, we shall present edge LAM for air-interface design in Section \ref{sec: air-interface}, with specific emphasis on channel prediction and beamforming design. 
% This requires the edge LAM to be capable of capturing the key characteristics of wireless data, adapting to time-varying channel condition and network topology, as well as making low-latency decisions for various tasks
% \begin{figure*}[t] 
% 	\centering
% 	\includegraphics[width=\linewidth]{img/edgeLAMapp.pdf}
% 	\caption{Edge LAM enabled intelligent services: a) vehicular edge LAM; b) robotic edge LAM.}
% 	\label{fig: application}
%         \vspace{-0.5cm}
% \end{figure*}

{\color{black}Considering the promising applications of edge LAMs for future wireless communication systems, this article aims to provide a comprehensive framework for the deployment of edge LAMs and their applications in air-interface designs from the perspective of model decomposition and resource allocation.
\begin{enumerate}
	\item [$\bullet$]In Section \ref{sec: training}, we present a collaborative training framework tailored for edge networks, which encompasses both federated fine-tuning (FedFT) and full-parameter parallel training, incorporating novel model decomposition, resource allocation, and synchronization schemes to support efficient on-device training at the network edge\cite{10855336}.
	\item [$\bullet$]In Section \ref{sec: inference}, we present a microservice-based edge LAM inference framework that decomposes LAM into functional modules deployed at edge devices, reorganizing multimodal downstream inference services as dynamic microservice flows. 
    We further investigate microservice deployment, robust orchestration, and online migration for edge LAM inference, leveraging real-time learning and optimization techniques to enhance performance.
	\item [$\bullet$]In Section \ref{sec: air-interface}, we introduce edge LAM for air-interface design, with a particular focus on channel prediction and beamforming design. We detail the design principles that enable edge LAM to effectively capture the essential characteristics of wireless data, adapt to dynamic channel conditions and network topology changes, and make low-latency decisions across diverse tasks. Additionally, a concise case study is presented in Section \ref{sec: casestudy} to demonstrate the effectiveness of the proposed framework in applying federated LAM to channel prediction.
\end{enumerate}}

\section{Collaborative Training of Edge LAMs}\label{sec: training}

Training LAMs to uncover latent patterns is crucial for enabling various intelligent applications, which, however, requires significant resources and data consumption. 
To address this issue, parallel training that leverages distributed resources and data can accelerate the training process. Nonetheless, existing distributed training frameworks heavily rely on cloud computing and require collecting large-scale data for training. This incurs overwhelming communication overhead and raises severe privacy concerns.
These issues can be mitigated through on-device training, which utilizes local computing resources to perform training while ensuring that raw data remain in situ \cite{9606720}. This approach minimizes data transmission to cloud servers and reduces communication overhead. However, the limited data, computation, and storage resources of edge devices restrict the direct application of on-device training to LAM training.
To fully harness the representational power of LAMs while minimizing communication costs and ensuring data privacy, we propose a collaborative training framework that facilitates the fine-tuning of pre-trained LAMs and the full-parameter training of lightweight LAMs.
% \begin{figure*}[t] 
% 	\centering
% 	\includegraphics[width=0.8\linewidth]{img/FedFT_v2.pdf}
% 	\caption{{\color{black}Federated fine-tuning over wireless networks. The left side of the figure demonstrates the communication architecture of the proposed FedFT framework; the right side of the figure shows the simulation results, analyzing training loss and test accuracy under varying average transmission latency constraints. Specifically, under stringent transmission latency constraints, the edge server jointly optimizes device scheduling and bandwidth allocation to enhance learning performance. Two benchmark methods are considered: GS (Greedy Strategy) and AABA (Adaptive Average Bandwidth Allocation). The proposed algorithm leverages Lyapunov theory to optimize radio resources and device scheduling from a long-term perspective. This approach aligns with the fact that the fine-tuning performance of LAMs is significantly influenced by multiple rounds of communication, achieving superior learning performance in comparison to the benchmarks.}}
% 	\label{fig: fedft}
%     \vspace{-0.5cm}
% \end{figure*}
\begin{figure}[t] 
	\centering
	\includegraphics[width=\linewidth]{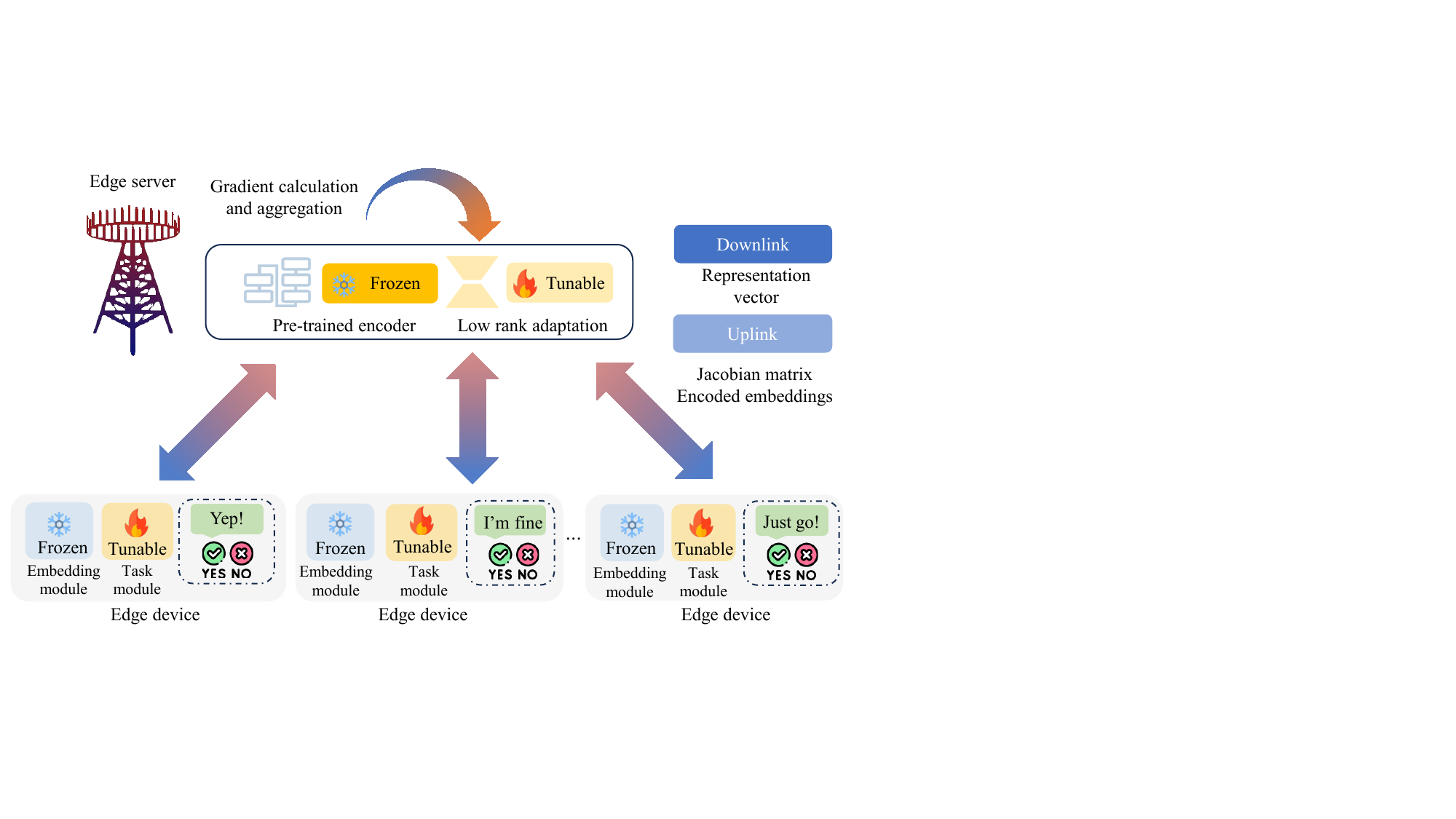}
	\caption{Federated fine-tuning over wireless networks.}
	\label{fig: fedft}
    \vspace{-0.5cm}
\end{figure}

\subsection{Federated Fine-Tuning for Edge LAMs}
Supporting personalized LAM services requires the fine-tuning of pre-trained LAMs with task-specific datasets, which, however, are typically isolated at different edge devices and highly sensitive.
To address these issues, we propose a FedFT framework, as shown in Fig. \ref{fig: fedft}, which enables the collaborative tuning of a global foundation model by coordinating distributed edge devices \cite{10855336} in a privacy-preserving manner.
However, it is impractical to deploy a full pre-trained LAM on each edge device with limited storage and computation capabilities.
This issue can be addressed by constructing a split FedFT framework, which decomposes the LAMs into the embedding module, decoder module, and task module, where the lightweight embedding and task modules are deployed at edge devices while keeping the computationally expensive decoder module at the edge server. 
To reduce training latency due to sequential communication, we propose a low-rank adaptation (LoRA) based FedFT framework to decompose the trainable parameters of the decoder into a series of low-rank matrices while freezing the original parameters. 
{\color{black}Thus, the edge server processes the received embeddings in parallel with the added low-rank matrices and the original encoder to form representation vectors. Meanwhile, the low-rank matrices are updated using Jacobian vectors aggregated from edge devices in the backward propagation. These Jacobian vectors result in significantly lower transmission overhead compared to the conventional FL framework.}
The proposed FedFL framework differs from the conventional FL framework in two key aspects. First, the edge server in the proposed framework transmits representation vectors to edge devices using unicast transmission instead of broadcast transmission in the conventional FL framework. 
Second, the edge server in the proposed framework performs federated averaging over the gradient of low-rank matrices instead of the full model in the conventional FL framework. 
By folding the insights obtained from the theoretical analysis into the transmission scheme design, we develop an online resource allocation algorithm to enhance the learning performance \cite{10855336}. 

Although FedFT protects raw data privacy by aggregating the gradients of the loss w.r.t representation vectors at the edge server, these gradients remain vulnerable to inference attacks, such as gradient inversion by untrusted servers. 
To address this issue, we enhance the FedFT system by integrating differential privacy (DP), where the edge server aggregates noisy gradients of the loss to ensure sample-level privacy protection.
Since the analytical characterization of DP within the FedFT context is implicit, we introduce a Rényi DP (RDP)-based FedFT framework, which leverages the divergence between probability density functions of transmitted gradients, computed locally at edge devices, to determine the optimal noise level. 
{\color{black}Unlike conventional FL systems with DP, where the server aggregates noisy gradients of the entire model, the FedFT system introduces cascading noise through the chain rule, as gradients of the loss are multiplied with those of low-rank matrices. 
This difference may hinder the convergence of the considered FedFT system. To address this, theoretical analysis shall be conducted to evaluate the impact of added noise on the convergence behavior and privacy guarantees. Then, the trade-off between privacy and learning performance can be balanced by adaptively tuning noise injection based on channel conditions.
}

% \begin{figure}[!] 
% 	\centering
% 	\includegraphics[width=\linewidth]{img/parallelism.pdf}
% 	\caption{Looped tensor parallelism for full-parameter edge LAM training.}
% 	\label{fig: loop}
%         \vspace{-0.5cm}
% \end{figure}
\subsection{Looped Tensor Parallelism for Full-Parameter Edge LAM Training}
%逻辑
Mainstream LAMs are essentially transformer-based deep neural networks (DNNs) with massive neurons, which are trained with massive data for contextual modeling. 
Unlike conventional learning tasks (e.g., ResNet18), which can be performed on edge devices with limited resources (e.g., NVIDIA Nano with 4 GB RAM), full-parameter training of lightweight LAMs, such as BERT, typically requires at least 10 GB of RAM per batch, far exceeding the computational and storage capabilities of a single device. 
This necessitates the use of parallel training mechanisms (e.g., pipeline and data parallelism), which, however, are typically designed for centralized cloud training \cite{deng2024cloud}. 
To leverage distributed resources for the training of lightweight LAM, we propose a looped tensor parallelism framework, which employs a consistent set of network parameters to represent computationally intensive transformer blocks, and decomposes network parameters into distributed matrix computations on edge devices.
{\color{black}Since tensor-based general matrix multiplication dominates the computation burden in the full-parameter training of lightweight LAMs, tensor parallelism can decompose large-scale matrix calculations in the attention and feed-forward networks of the transformer into multiple small-scale matrix calculations. These smaller tasks are then performed by different edge devices, thereby relieving computation overhead.}

To circumvent raw data sharing, raw data can be encoded by split weight matrices at edge devices, which are then aggregated by the edge server for nonlinear activation. 
This is achieved by adopting the transposed matrices in computation and merging the intermediate activations at the edge server.
{\color{black}To mitigate the storage constraints of edge devices, a cross-layer parameter sharing technique can be applied to split the attention and feed-forward networks, where the cascaded transformer blocks at different depths reuse the same parameters in a looped manner. }
% , as shown in Fig. \ref{fig: loop}
The innovative workflow results in new challenges in matching the edge resources and the training latency. 
Specifically, the training latency is influenced by the heterogeneous computation capabilities of edge devices, time-varying channel conditions, and limited radio resources. 
Besides, the requirement for tensor merging imposes the need to maintain consistent matrix dimensions across devices, constraining task sizes to the minimum available storage of the participating edge devices.
Hence, minimizing the training latency requires the joint optimization of communication, computation, and storage resources. 

\begin{figure*}[t]
	\centering
	\includegraphics[width=\linewidth]{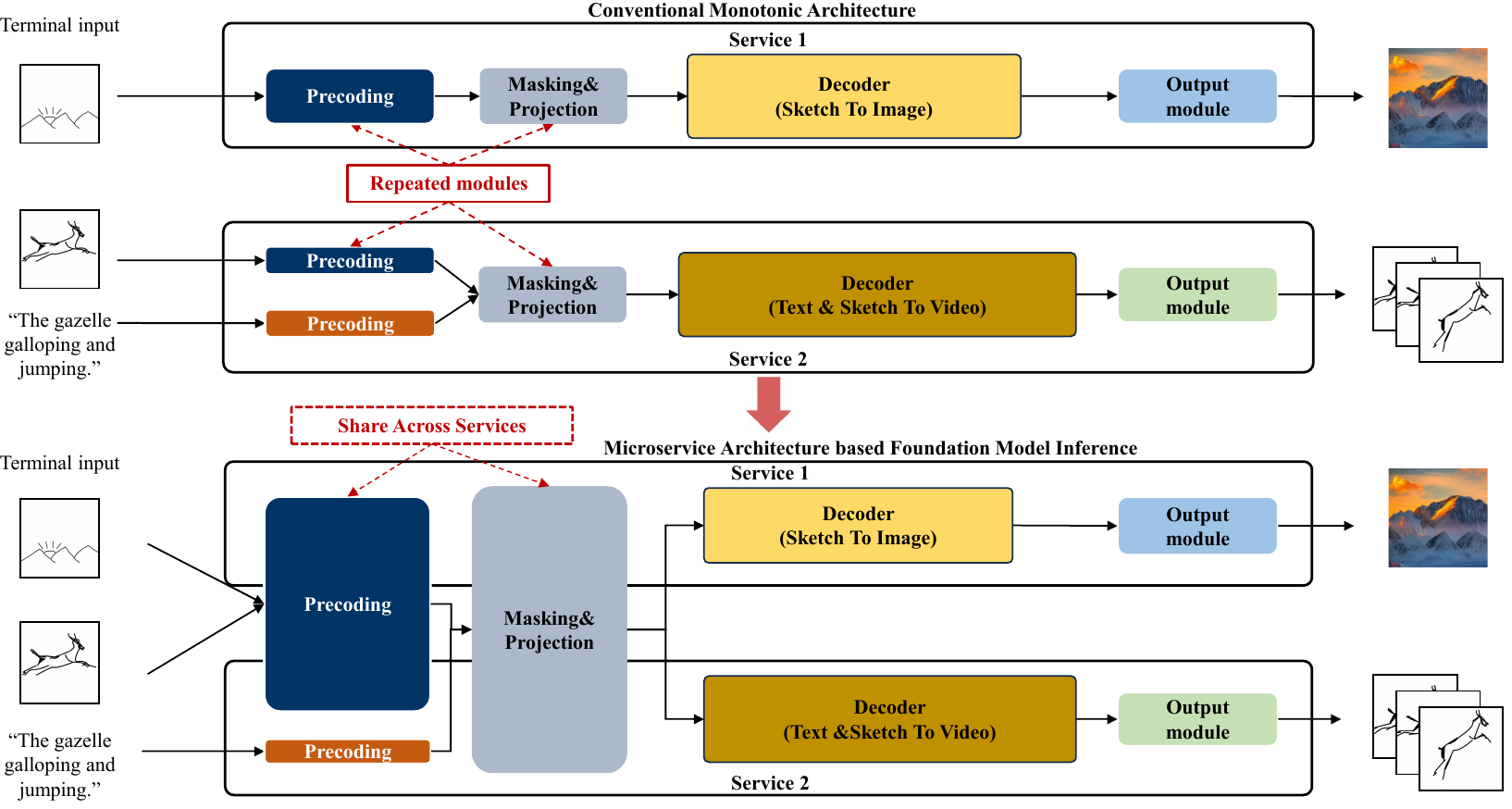}
	\caption{An illustration of the proposed microservice architecture for edge LAM.}
	\label{Fig::overview_microservice}
        \vspace{-0.5cm}
\end{figure*}

\section{Microservice-Enabled Multimodal Edge LAM Inference}\label{sec: inference}
% Unlike traditional edge AI inference with standalone neural networks, edge LAM inference operates as a sequential process involving multiple specialized neural networks. Each functional module within a LAM performs a distinct task, such as pre-coding or denoising, working collaboratively to handle diverse user requests.
% While LAMs excel in providing versatile services, they introduce challenges distinct from conventional AI models. 
% For instance, traditional AI models (e.g., AlexNet) are compact and simply split to be deployed on resource-constrained edge devices, while LAMs are significantly larger in size and require the orchestration of multiple interconnected modules.
% Another challenge is that the successive processing nature of LAMs poses a significant hurdle to real-time inference.
{\color{black}
Unlike traditional edge AI inference with standalone neural networks, multimodal edge LAM inference operates as a sequential process involving multiple specialized module and each performs a distinct task, such as pre-coding or denoising. However, traditional monolithic approaches for deploying them result in redundant deployment and calculation, reducing resource utilization\cite{10591707}. 
% Moreover, traditional monolithic approaches for deploying LAM-enabled services often result in redundant functional modules, leading to low resource utilization\cite{10591707}.
To address these issues, we propose a microservice-based approach that decomposes LAM services into functional components, supporting diverse multimodal downstream inference tasks by disassembling large models into lightweight microservices and reusing shared components, as shown in Fig. \ref{Fig::overview_microservice}. }
Given that both resource availability and user requests are stochastic, we further investigate deployment strategy, robust orchestration, and online migration via real-time status learning and decision making.

\subsection{Microservice Deployment for Edge LAM Inference}
Mainstream LAMs comprise modality encoders, input projectors, a backbone calculator, an output projector, and modality decoders, facilitating various inference tasks through the sequential processing of requested data via these modules. However, the modalities supported by existing single LAMs are limited. For example, DALL-E accommodates only text and image modalities, whereas VideoGPT supports video and text modalities.
To meet the complex modal requirements, different LAMs need to be deployed on edge devices, resulting in the repeated deployment of projectors and modality encoders. For instance, the speech and vision transformer uses the same vision encoders as VisualBERT while incorporating a unique speech encoder. This deployment leads to the duplication of the vision encoder at the edge devices, introducing redundancy that results in low resource utilization.
{To enhance resource utilization, we propose a microservices-based inference framework to virtualize the functional modules (e.g., modality encoders, input projectors) of multimodal LAMs into a series of independent microservices, each tailored to perform a specific function. This architecture transforms the process into a unidirectional acyclic graph consisting of distinct microservices, where nodes represent different microservices and edges denote the data dependencies between them. This design transforms inference tasks based on different multimodal LAMs into distinct microservice flows shared among edge devices, which can be further optimized to enhance overall efficiency, scalability, and responsiveness to dynamic changes in user requests.
}

Given the statistical requirements of inference tasks, microservice deployment emerges as a key concern \cite{yang2024latency}, which refers to the strategic allocation of computation and communication tasks to address the mismatch between the stringent latency requirements and inference tasks.
Its goal is to minimize total latency while efficiently utilizing limited resources across heterogeneous edge devices. Thus, the microservice deployment problem can be formulated as an NP-hard combinatorial optimization problem concerning the discrete-valued deployment indicator.
{To efficiently tackle this combinatorial optimization problem, training a likelihood model that approximates the desired distribution of the near-optimal deployment strategy is a promising approach. This model allows for real-time deployment indicator generation based on the status of edge devices in an end-to-end manner. However, constructing the likelihood model using the marginal sample probability of a latent variable model introduces an intractable upper bound due to the latent variables.
To address this issue, the posterior probability w.r.t the latent variable can be mapped onto a pre-defined stationary distribution, which can be learned via diffusion models in a data-driven manner.}
\subsection{Microservice Orchestration and Migration}
User mobility determines the utility of deployed microservice systems. 
The uncertainty of user requests can also affect the system latency and operational costs of the deployed microservices, thereby reducing the reliability of edge LAM inference and underscoring the urgent need for reliable microservice orchestration.
% For instance, when a moving car requests inference services from a roadside unit, the inference latency varies as the car's position changes. While at congested intersections, transient high-frequency user requests can consume communication and computational resources between roadside units, thereby increasing inference latency.
% This underscores the urgent need for reliable microservice orchestration.
Reliable microservice orchestration in edge LAM inference is defined as the coordination of computation and communication tasks related to designing a robust service routing scheme for user requests.
Different from microservice deployment, microservice orchestration assumes that the same microservice is deployed on at least two or more different edge devices to respond to user requests, where the goal is to find a robust routing strategy that maximizes the long-term system utility under dynamic user requests.
This can be viewed as a Markov decision process, where the reward function is uncertain due to user mobility, and determined by the network status and routing designs in previous decision rounds. Thus, robust adversarial reinforcement learning can be adopted to establish the desired orchestration policy.
This is achieved by utilizing a primary agent to learn orchestration strategies that maximize long-term system utility under dynamic requests and mobilities, while an adversarial agent generates request strategies aimed at challenging the primary agent with minimized utility. 
The desired orchestration policies can be learned by parameterized DNNs in a data-driven manner.

Microservice migration is another way to address the user mobility issue. 
Specifically, the limited coverage area of edge servers, coupled with high mobility scenarios (e.g., utilizing edge LAM inference services on highways or high-speed trains), may result in significantly high inference latency. 
This dynamic fluctuation in resource availability and service requests requires the migration of both microservices and user data from the previously serving edge server to a more suitable one. 
Due to the spatial and temporal correlations of user locations, the inference latency is directly determined by the microservice migration decisions. 
To maximize the time-average expectation of system utility, the microservice migration problem is a long-term optimization concerning migration costs (e.g., bandwidth, storage, and power), while accounting for latency, jitter, and packet loss requirements. 
By employing Lyapunov optimization, the goal of meeting long-term constraints can be transformed into a virtual queue to indicate the gap between the resource utility and the budget.
Inspired by the drift-plus-penalty algorithm of Lyapunov optimization, the long-term optimization problem can be transformed into multiple one-shot migration decision-making problems. 
This microservice migration cost minimization problem can be solved by developing either an alternating optimization algorithm or an online learning algorithm to dynamically allocate computation and communication resources.

\section{Edge LAMs for Air-interface Design}\label{sec: air-interface}

{\color{black}Air-interface technologies have long served as the foundation of wireless communication systems. While conventional optimization-based methods leverage domain expert knowledge to achieve desirable network performance, they struggle with scalability in large-scale networks.}
By leveraging DNNs to directly establish the nonlinear mapping between system states and optimization parameters, AI-based methods can solve diverse network optimization and management problems. 
However, they are typically designed to be scenario- and problem-specific, suffering from poor generalization capability. 
{\color{black}In particular, frequent retraining is incurred to adapt to rapidly changing network conditions, posing significant challenges to efficiency and scalability. }
In contrast, edge LAMs excel at extracting complex spatio-temporal features of training data and learning in-context embeddings, demonstrating enhanced generalization capabilities. 
Training edge LAM with large-scale wireless data enables it to function as a universal feature extractor, effectively capturing intricate patterns and representations that can be applied to channel prediction and beamforming \cite{10579546,10716720}. 
{\color{black}To unleash the full potential of edge LAM for air-interface, it is essential to investigate the precoder design for effective processing of wireless data, which is a unique modality distinct from texts and images, task-specific generator design to resolve domain-specific constraints and ever-changing network dimensions, and fine-tuning with the domain knowledge from the wireless environment.}

\subsection{Federated LAM for Channel Prediction}
Accurate channel prediction based on historical channel state information (CSI) data is vital for achieving high communication performance in highly dynamic 6G networks, where the channel estimation overhead can be huge because of the short channel coherence time. 
Facing the difficulty of accurately characterizing practical wireless channels, the conventional sparsity-, prony-, and extrapolation-based methods suffer from limited prediction performance. 
By learning the temporal and frequency channel variations in a data-driven manner, AI-based methods can support channel prediction in both single- and multi-antenna systems. However, existing AI-based methods suffer from the limitations of poor channel feature extraction capability because of the small-sized neural networks and weak generalization capability because of the scenario-specific design. 
{\color{black}Empowered with massive parameter counts and the transformer architecture, edge LAM is able to model complex time-varying wireless networks with in-context generalization, instead of constructing scenario-specific feature-prediction mappings. Consequently, pre-training LAM with massive CSI datasets can support multiple channel prediction tasks, without the need to deploy multiple configuration-specific models \cite{liu2024wifowirelessfoundationmodel}. }
However, diverse three-dimensional CSI datasets containing massive training samples should be collected and transmitted to a central server for pre-training, incurring excessive communication overhead and severe privacy concerns. 

Federated LAM holds the potential to enable effective channel prediction while facilitating communication-efficient and privacy-preserving model training by aggregating only the gradients of trainable parameters. 
Specifically, edge devices situated in different locations perceive varying channel samples under different channel distributions and network configurations, enriching data diversity and improving model generalization and adaptability of federated LAM. 
{\color{black}To capture the spatial, temporal, and frequency correlation between successive channel states, an auto-regressive neural network-based encoder can be adopted to model the channel dependency, extracting embedding features and mapping them into token representations for federated LAMs. }
Moreover, low-rank matrix decomposition methods, such as LoRA, can be employed in federated LAM adaptation, given their capabilities in achieving learning performance comparable to full-parameter tuning with only 5\% of the parameters, significantly reducing both computational costs and communication overhead. 
By further specifically designing pre-processing modules for CSI data, a well-trained model via federated LAM can effectively support channel prediction across diverse scenarios and configurations. 

\subsection{Graph LAMs for Beamforming}
Beamforming design is crucial for enhancing the spectrum- and energy-efficiency of multi-antenna systems, facilitating directional communication, sensing, and positioning. 
To meet the ever-growing demands for high data-rate and low-latency communications, many advanced multi-antenna technologies, e.g., extremely large MIMO, reconfigurable intelligent surface, and holographic MIMO, have emerged to provide unprecedented capacity, which, however, poses significant challenges to beamforming design.
Because of the high-dimensional nature and often coupled optimization variables, beamforming design problems are usually non-convex, which cannot be efficiently tackled by optimization-based methods. 
Integrating AI into beamforming design holds great promise, yet it also encounters several critical challenges. 
First, with constrained model size, existing AI-based methods cannot sufficiently characterize the channel model, limiting their ability to effectively map wireless parameters to optimal beamforming design \cite{sheng2024Beam}.
Second, because of the problem-specific model training, existing AI-based methods exhibit weak generalization capability and cannot simultaneously tackle multiple tasks. 
Third, without fully exploiting the underlying network topology and the permutation properties of the beamforming problem, existing AI-based methods suffer from poor scalability and achieve much degraded performance as the network size changes. 

Thus, we propose a graph LAM-based beamforming framework, which can leverage massive neurons to enable rich and contextualized channel feature extraction, utilize high-quality datasets spanning diverse channel environments to ensure strong generalization ability, and exploit network topology and permutation properties to achieve exceptional scalability. 
Specifically, beamforming design for various scenarios (e.g., hybrid beamforming, joint active and passive beamforming) can be formulated as graph optimization problems, where the distributed communication devices (e.g., edge devices, base stations, and reconfigurable intelligent surface) are represented as nodes, and the mutual communication and/or sensing dependencies are represented as edges. 
To bridge the gap between non-Euclidean graph data with topological information and Euclidean tokens suitable for pre-trained LAMs, a graph neural network precoder can be adopted to extract hidden features, followed by a lightweight neural network that projects these features into the token space. 
Beyond adopting low-rank matrix decomposition for wireless-specific adaptation, projection and normalization modules can ensure that the generated beamforming design satisfies domain-specific constraints, e.g., transmission rate, sensing resolution, and learning performance. 
Moreover, end-to-end design that bypasses explicit channel estimation and directly utilizes noisy pilots can be further integrated into the graph LAM-based beamforming design framework, thereby mitigating the performance degradation due to the objective mismatch between the beamforming design and channel estimation modules. This can be achieved by developing a complex-valued neural network for encoding pilot sequences into tokens. 
{\color{black}To further mitigate the communication overhead caused by the collection of local CSI and the dissemination of beamforming vectors for centralized learning, FedFT can be used to enable distributed and collaborative learning under different network architectures and configurations.}
{\color{black}
\section{Case Study}\label{sec: casestudy}
This subsection evaluates the learning performance of the proposed federated LAM for channel prediction.
% \begin{figure}
%     \centering
%     \subfigure[Convergence Performance]
% 	{
% 			\includegraphics[width=0.85\linewidth]{img/FedFT4CPTDD.eps}
% 			\label{fig1_1}
% 	}
%     \\
% 	\subfigure[Average Inference Time]
% 	{
% 			\includegraphics[width=0.85\linewidth]{img/inferencetime.eps}
% 			\label{fig1_2}
%             }
%     \caption{{\color{black}Results of federated LAM for channel prediction.}}
%     \label{fig:app_fedft4cp}
% \end{figure}
\begin{figure}
    \centering
    \includegraphics[width=\linewidth]{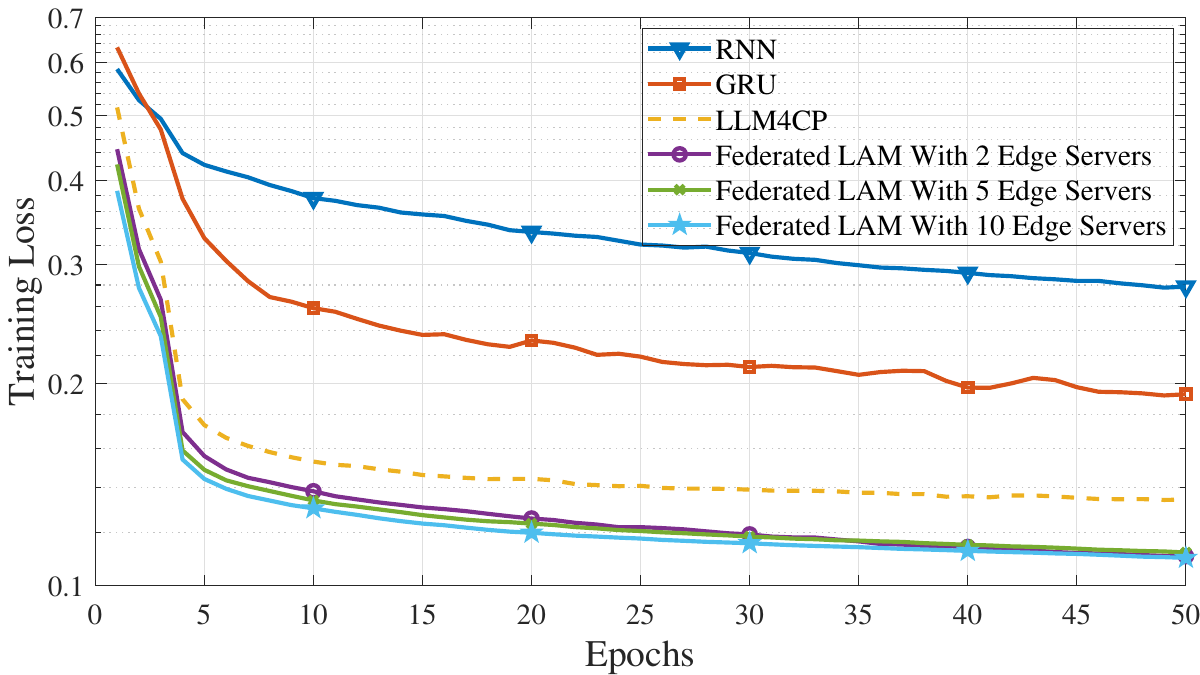}
    \caption{{\color{black}Federated LAM for channel prediction.}}
    \label{fig:app_fedft4cp}
    \vspace{-0.5cm}
\end{figure}

\textbf{Important Settings:} We employ the LLM-empowered channel prediction method (LLM4CP) as the base model \cite{liu2024wifowirelessfoundationmodel} and utilize QuaDRiGa to generate 3GPP-compliant channel realizations as the training dataset. We consider up to 10 edge servers, each equipped with 5\% independent and identically distributed channel samples. 

\textbf{Evaluation Results:} The performance of the proposed method is illustrated in Fig. \ref{fig:app_fedft4cp}. For comparison, we consider three baselines: fine-tuning LLM4CP, gated recurrent unit (GRU), and recurrent neural network (RNN) using only local data with a stochastic gradient descent (SGD) optimizer. Federated LAM achieves lower training loss than these benchmarks. Its convergence performance initially improves with more devices but stabilizes over time. By outperforming local training, federated LAM enables efficient, privacy-preserving channel prediction for air-interface design.
}
\section{Conclusions}
In this article, we have explored the potential of edge LAMs as a transformative technology for achieving AI-native 6G networks. Our investigation focused on two key perspectives: ``model decomposition'' and ``resource management'', which are crucial for addressing the challenges of deploying and utilizing LAMs at the network edge. 
We proposed a collaborative training framework that leverages split FL and looped tensor parallelism to support communication-efficient and trustworthy fine-tuning and full-parameter training of edge LAMs. To provide low-latency inference services, we developed a microservice-enabled architecture that dynamically matches the computation demands of microservices with the capabilities of edge devices. Furthermore, we investigated applications of edge LAMs for channel prediction and beamforming. Despite these advancements, edge LAMs face significant challenges, including model size constraints due to limited storage and computation resources on edge devices, high energy consumption that may deplete battery life quickly, and privacy concerns arising from potential data leakage during model training and inference. Addressing these challenges will require innovative approaches to model compression, energy-efficient computing, and robust privacy-preserving techniques. 
% We hope this article serves as a valuable reference for envisioning edge LAMs in the context of native AI-enabled 6G, sparking further research endeavors in this exciting and promising field. 

\bibliographystyle{IEEEtran}
\bibliography{ref.bib}

\vspace{-6mm}
\begin{IEEEbiographynophoto}{Zixin Wang} (eewangzx@ust.hk) received his Ph.D. degree from the University of Chinese Academy of Sciences. He is currently a Postdoctoral Fellow with The Hong Kong University of Science and Technology. 
\end{IEEEbiographynophoto}
% \vspace{-6mm}
% \begin{IEEEbiographynophoto}{Yuanming Shi} (shiym@shanghaitech.edu.cn) received his Ph.D. degree from The Hong Kong University of Science and Technology. He is currently a Full Professor with ShanghaiTech University. He was a recipient of the 2016 IEEE Marconi Prize Paper Award in Wireless Communications, the 2016 Young Author Best Paper Award by the IEEE Signal Processing Society, and the 2021 IEEE ComSoc Asia-Pacific Outstanding Young Researcher Award. He is an IET fellow.
% \end{IEEEbiographynophoto}
\vspace{-6mm}
\begin{IEEEbiographynophoto}{Yuanming Shi} (shiym@shanghaitech.edu.cn) received his Ph.D. degree from The Hong Kong University of Science and Technology. He is currently a Full Professor with ShanghaiTech University and an IET fellow.
\end{IEEEbiographynophoto}
\vspace{-6mm}
\begin{IEEEbiographynophoto}{Yong Zhou} (zhouyong@shanghaitech.edu.cn) received his Ph.D. degree from University of Waterloo. He is currently an Associate Professor with ShanghaiTech University. 
\end{IEEEbiographynophoto}
\vspace{-6mm}
\begin{IEEEbiographynophoto}{Jingyang Zhu} (zhujy2@shanghaitech.edu.cn) is pursuing his Ph.D. degree with ShanghaiTech University. 
\end{IEEEbiographynophoto}
\vspace{-6mm}
\begin{IEEEbiographynophoto}{Khaled B. Letaief} (eekhaled@ust.hk) received his Ph.D. from Purdue University. He has been with HKUST since 1993, where he was an acting provost and dean of engineering. He is now a Senior Advisor to the President and the New Bright Professor of Engineering. From 2015 to 2018, he joined HBKU in Qatar as Provost. He is an ISI Highly Cited Researcher and a recipient of many distinguished awards. He has served in many IEEE leadership positions, including ComSoc president, vice-president for technical activities, and vice-president for conferences.
He is a member of the US National Academy of Engineering. 
\end{IEEEbiographynophoto}

\end{document}